\documentclass[10pt, aps,twocolumn,amsmath,amssymb]{revtex4-1} 
\pdfoutput=1
\usepackage{graphicx,epsfig}
\usepackage{natbib}
\usepackage{bm} 
\usepackage{color}
\usepackage{multirow}
\usepackage{amsfonts,amsthm,dsfont}
\usepackage{wasysym}
\newcommand{\tgreen}[2]{
\left<\hspace{-2pt}\left<#1,#2\right>\hspace{-2.3pt}\right>_{\hspace{-2pt}t}^{\hspace{-2pt}r}
}

\newcommand{\wgreen}[2]{
\left<\hspace{-2.3pt}\left<#1,#2\right>\hspace{-2.3pt}\right>_{\hspace{-2pt}\varepsilon}^{\hspace{-2pt}r}
}

\newcommand{\pa}[1]{\left ( #1 \right )}

\begin{document}
\title{Finite $U$ thermoelectrical transport in graphene based quantum dots}

\author{Jos\'{e} Ram\'{o}n Isern-Lozano}
\affiliation{Instituto de F\'{\i}sica Interdisciplinar y Sistemas Complejos IFISC (UIB-CSIC), 
             E-07122 Palma de Mallorca, Spain}
\author{Ioan Grosu}
\affiliation{Department of Theoretical Physics, University of Cluj,
400084 Cluj-Napoca, Romania}
\author{Rosa L\'opez}
\affiliation{Instituto de F\'{\i}sica Interdisciplinar y Sistemas Complejos IFISC (UIB-CSIC), 
             E-07122 Palma de Mallorca, Spain}
\affiliation{Kavli Institute for Theoretical Physics, University of California, Santa Barbara, 
             California 93106-4030, USA}
\author{Jong Soo Lim}
 \affiliation{School of Physics, Korea Institute for Advanced Study, Seoul 130-722, Korea}
\author{Mircea Crisan}
\affiliation{Department of Theoretical Physics, University of Cluj, 400084 Cluj-Napoca, Romania}

\begin{abstract}
We study the thermoelectrical transports for an interacting dot attached to two graphene electrodes. 
Graphene band structure shows a pseudogap density of states that affects strongly the transport properties. 
In this work, we focus on the Coulomb blockade regime and derive the expression for Onsager matrix $\{\mathcal{O}_{ij}\}$ that relates the electrical and heat currents with electrical and thermal biases in the linear response regime. 
Our findings show double peak structures for the electrical and thermal conductances versus the dot level in accordance with the Coulom blockade phenomenon. 
Remarkably, however, the thermal conductance is much smaller than the electrical conductance, resulting in high figure of merit value for some gate voltage.
Finally, we report a large departure from the Wiedemann-Franz law caused mainly by the pseudogap density of states in the contacts and weakly affected by interactions.
\end{abstract}


\maketitle

\section{Introduction}
Graphene, a mono-layer of carbon atoms disposed on an hexagonal lattice was firstly  synthesized a decade ago \cite{RevModPhys.81.109, Geim_2007}. However, graphene was not an unknown material, but it was theoretically 
investigated long time ago by P. R. Wallance and others 
\cite{Wallace47,Lomer1955,McClure1956,McClure1957,Slonczewski1958,Bassani1967} as a two-dimensional graphite 
material. The unusual properties of graphene are owing to its Dirac-like band structure, where conduction and 
valence bands touch at six discrete points at the edges of the honeycomb Brillouin zone. The relativistic (Dirac) 
character of graphene was pointed out by D.P. DiVincenzo and E.J. Mele \cite{DiVincenzo1984} before it was 
created in the lab.  In graphene band structure, only two of the six Dirac cones are nonequivalent being 
currently named as K and K' points \cite{RevModPhys.81.109}. The quasiparticle excitations at those 
points obey linear Dirac-like energy dispersion and are responsible for many physical phenomena such as the 
relativistic quantum Hall, the Klein tunneling  among others \cite{RevModPhys.81.109}. Besides, graphene 
quasiparticles are chiral fermions with potential applications in the so-called valleytronics \cite{Nebel_2013,Lundeberg_2014}.
\begin{figure}
\centering
\includegraphics[width=.4\textwidth,keepaspectratio=true]{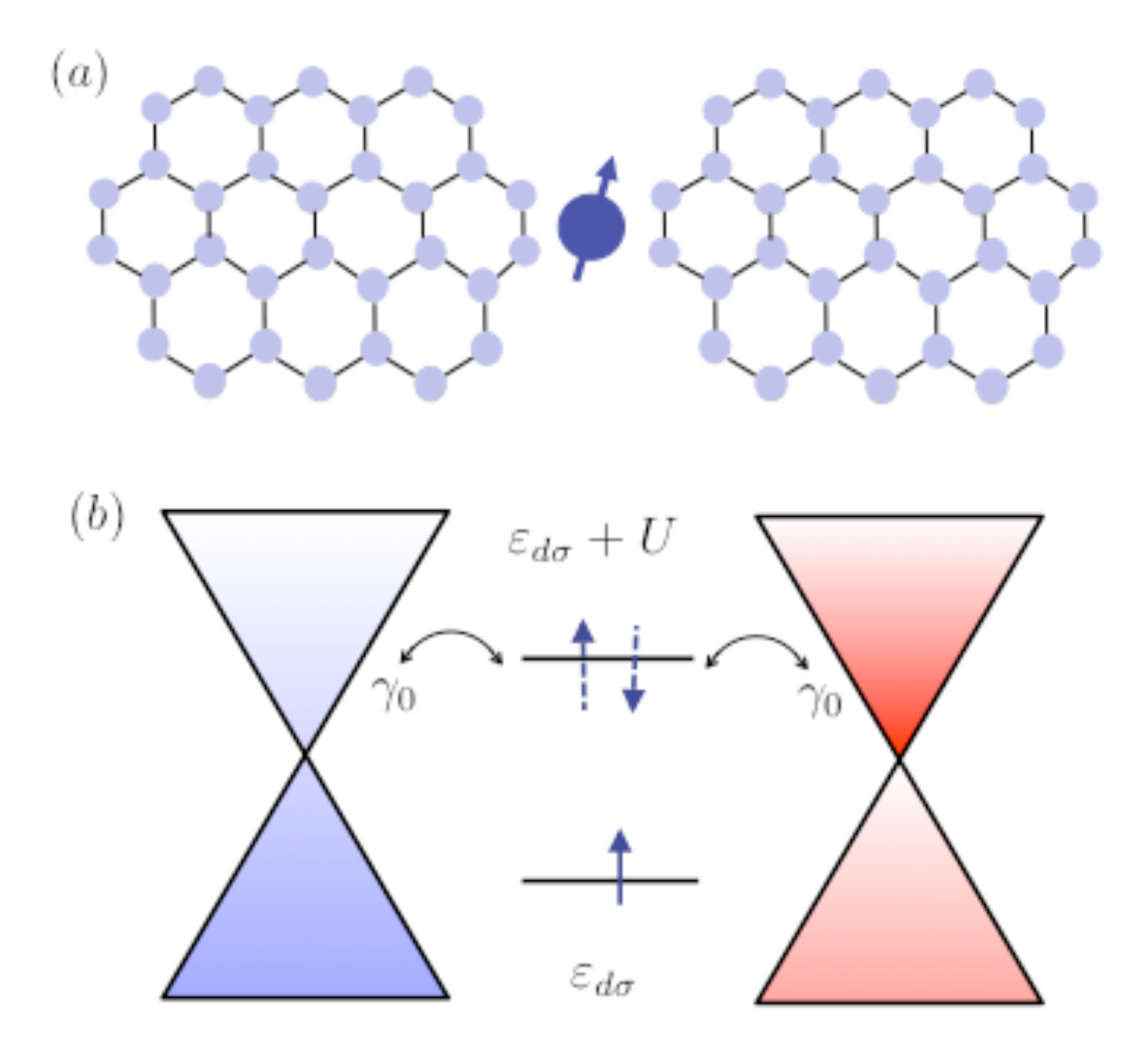}
\caption{ (a) Two graphene flakes attached to a confined interacting site. Different temperatures and electrical potentials are applied to the left and right graphene electrodes. 
(b) Energy band diagram for the system under consideration. The dot level is denoted by $\varepsilon_{d\sigma}$. For double occupancy the dot level is shifted by an amount $U$ that describes the Coulomb repulsion. 
The hybridization $\Gamma(\varepsilon)=\gamma_0 |\varepsilon-\mu|$ depends on energy $\varepsilon$. We consider undoped graphene electrodes with $\mu=\varepsilon_F=0$ located at the Dirac point.}
\label{fig:1}
\end{figure}
One of the successful graphene applications is found when graphene acts as a conductive material attached to a 
nanostructure to create a single-electron transistor circuitry \cite{Stampfer_2008,Tan_2011}. An instance is the 
particular case of a quantum dot carved in a graphene sheet \cite{Sols_2007,Stampfer_2008,Tan_2011,Murali_2012}. Here, typical Coulomb blockade phenomenon is 
observed associated to the transport of charges across the localized dot level \cite{Stampfer_2008,Tan_2011}.
The linear conductance exhibits peaks at the dot resonances separated by the mean-dot level spacing and electron-electron Coulomb repulsion. 
Interestingly, transport of charges in graphene based 
quantum dots occur at much higher temperatures than in traditional semiconductor dots \cite{Murali_2012}. Besides, under certain circumstances many-body effects, such as the Kondo effect \cite{9780511470752} can be 
observed \cite{Cronenwett24071998,Goldhaber98}. In reality graphene alters quite strongly Kondo physics \cite{Hentschel_2007,CastroNeto2009,Uchoa_2009, Vojta_2010, Zhu_2011,Sengupta_2011,Chao_2011, Kharitonov_2013,Mitchell_2013,Fritz_2013}.  
The  Kondo effect in graphene has been intensively discussed in connection with the  magnetic impurity problem in 
massless Dirac fermions. The experimental measurements show that the Kondo effect can be induced by lattice 
vacancies in graphene sheets \cite{Chao_2011,Kharitonov_2013} . The  Kondo  problem  in massless Dirac fermion 
system is indeed a particular case of the pseudo-gap Anderson model \cite{Glossop_2005} where the density of 
states (DOS) of conduction electrons is proportional to $|\varepsilon|^{r}$, with $\varepsilon$ being the energy of conduction 
electrons and $r$ an exponent specific for the material. The massless Dirac fermion system corresponds to $r=1$.  
There has been a great effort (theory and experiment) to describe the Kondo effect in Dirac-like materials. 
However, much less attention has received the Coulomb blockade regime in which charge fluctuations are the 
dominant events in transport. Coulomb-blockade effects are ubiquitous and govern the transport properties of a 
large variety of systems: quantum dots \cite{Nazarov_2009}, molecular bridges, carbon nanotubes \cite{Liang_2002}, 
etc. Coulomb blockade transport has been investigated extensively in nanostructures attached to metallic contacts. 
However, when the contacts are made by graphene the Coulomb blockade effect is an almost unexplored problem. The 
goal of this work is to investigate how Coulomb blockade effects are altered by the presence of Dirac-like 
fermions. Experimentally Coulomb blockade oscillations have been reported to occur in graphene quantum dots where 
the linear conductance oscillates with the dot gate position \cite{Stampfer_2008,Tan_2011}.

Graphene has been revealed as a singular material that exhibits amazing properties in a plenty of fields, namely, 
electronics, spintronics, optics, etc. \cite{RevModPhys.81.109}. A great portion of works has been devoted to the 
study of its purely electronic properties. In comparison, its thermal properties are poorly investigated. Recent 
measurements show that graphene offers high thermal conductance values \cite{Balandin_2011}. In this respect, 
graphene based materials offer new chances to the progress of the cross-field of thermoelectricity. Good 
thermoelectric devices at nanoscale are those where the heat dissipated or wasted is efficiently transformed into 
useful electricity and vice-versa \cite{Dubi2011}. The figure of merit $ZT$ is a coefficient that quantifies the 
efficiency in the heat-to-electricity (electricity-to-heat) conversion process. When the heat is only carried out 
by electrons then the $ZT$ is proportional to the square of the Seebeck (thermopower) coefficient or the ratio 
between the electrical and thermoelectrical linear conductances under the open-circuit condition.
 
The purpose of our work is to analyze the thermoelectric transport through a quantum dot (or interacting localized 
level) coupled to Dirac-like electrodes. We are interested in a temperature regime where charge 
fluctuations are important and therefore we discard the Kondo physics. Below we develop the theoretical 
model to describe this system. We will consider the simplest case where different voltages and temperatures are 
applied across the quantum dot. Then, pure electrical, thermal, and cross-conductance responses (thermoelectric)  
are characterized by the electrical ($\mathcal{G}$), thermal ($\mathcal{K}$),  and thermoelectrical 
($\mathcal{L}_{12}$) conductances, respectively. Our main findings indicate that graphene transistors offer 
extraordinary high thermoelectrical efficiencies. We report values of $ZT\approx 8$. Besides, due to the strong 
energy dependent tunneling rates, graphene interacting quantum dots do not follow the Wiedemann-Franz law.

The paper is organized as follows.  
In Sec.~II, our model for an interacting dot coupled to two Dirac-like contacts is introduced.
The Onsager matrix is computed to describe electrical and heat transports when 
electrical and thermal biases are applied, and the Seebeck and $ZT$ coefficients are defined in addition. 
Since the transport coefficients depend on the dot Green's function, 
it is furthermore calculated employing the equation-of-motion (EOM) technique \cite{Zubarev, Lacroix98, Meir1993, Kashcheyevs2006}. Our results for various transports are thoroughly explained in Sec.~III. 
Finally, Section~IV summarizes the main achievements of this work.

\section{Theoretical model}
\label{sec:theorical_model}
We employ an Anderson-like model to describe a spin 
degenerate localized level with strong on-site Coulomb interaction which is coupled to two Dirac-like electrodes as shown in Fig.~\ref{fig:1}. 
The full Hamiltonian consists of
$\mathcal{H}=\mathcal{H}_{D}+\mathcal{H}_{G}+\mathcal{H}_{V}$.
The localized level (dot) is described by
\begin{equation}
\label{eq:dot_hamiltonian} 
\mathcal{H}_{D} =\sum_{\sigma}  \varepsilon_{d\sigma} d_\sigma^\dagger d_\sigma + U n_{\uparrow} n_{\downarrow},
\end{equation}
where $d_{\sigma} (d_{\sigma}^{\dagger})$ annihilates (creates) an electron with spin $\sigma=\{\uparrow,\downarrow\}$ in the localized level,  
$\varepsilon_{d\sigma}$ denotes the spin-resolved energy level, and $U$ represents the strength of the Coulomb interaction. Here, 
$n_{\sigma} = d_{\sigma}^{\dagger}d_{\sigma}$ is the particle number operator for the localized level. 
The graphene contact Hamiltonian reads
\begin{equation}
\label{eq:leads_hamiltonian}
\mathcal{H}_{G} 
= \sum_{\alpha=L/R,s,k,\sigma} \int_{-k_c}^{+k_c} dk (\varepsilon_k - \mu)c_{\alpha sk\sigma}^{\dagger} c_{\alpha sk\sigma}
\end{equation}
where $\varepsilon_k = \hbar v_F k$ is relatively measured with respect to the chemical potential $\mu$ with $v_F\!\approx\! 10^{6}m/s$ being the graphene Fermi velocity. 
The chemical potential $\mu$ can be tuned by doping techniques and, hereafter, we consider $\mu=\varepsilon_F=0$. 
We recall that the linear dispersion $\varepsilon_k = \hbar v_Fk$ leads to the linear density of states 
\begin{equation}
\rho(\varepsilon) \propto |\varepsilon|
\end{equation}
The Hamiltonian then corresponds to the so-called Anderson pseudogap model \cite{Glossop_2003}.
Here, $c_{\alpha sk\sigma} (c_{\alpha sk\sigma}^{\dagger})$ annihilates (creates) a relativistic electron in the contact $\alpha$ 
($L/R$ for the left/right contact) with valley index $s$, wave-vector $k$, and spin $\sigma$. 
$k_c$ is a cutoff of the momentum such that $D =\pm \hbar v_F k_c$ is the cutoff of the energy. 
Dirac-like quasi-particles are 
tunnel coupled to the localized level by means of the hybridization Hamiltonian
\begin{equation}
\label{eq:hybridization_hamiltonian}
\mathcal{H}_{V} = \tilde{V} \sum_{i\,\sigma} \int_{-k_c}^{+k_c} dk \sqrt{|k|} \left(c_{\alpha sk\sigma}^\dagger d_\sigma + 
                    d_\sigma^\dagger c_{\alpha sk\sigma}\right),
\end{equation}
where  $\tilde{V} = V_0\sqrt{\pi\Omega_0}/2\pi$ being $\Omega_0$ the area of the graphene unit cell and $V_0$ the tunneling amplitude.

\subsection{Onsager matrix: transport coefficients}
In the next step, we compute the transport coefficients by employing the Onsager matrix \cite{onsager_1931}  that 
connects linearly charge and heat currents with the applied forces, which in our case are the electrical and 
thermal biases. 
We aim to investigate such transport coefficients in Dirac-like systems when Coulomb 
blockade phenomenon takes place. For such purpose we write down all the Onsager matrix elements in terms of the 
local density of states (DOS) of the localized impurity. Therefore, our description will depend on the 
approach employed to derive the impurity DOS. In the linear response regime, the Onsager matrix reads
\begin{equation}\label{eq_matrix}
\pa{
\begin{array}{c}
\bm{\mathcal{I}} \\ 
\bm{\mathcal{J}}
\end{array} }
=\pa{
\begin{array}{cc}
\mathcal{L}_{11} & \mathcal{L}_{12} \\ 
\mathcal{L}_{21} & \mathcal{L}_{22}
\end{array}}
\pa{
\begin{array}{c}
\Delta V \\
\Delta T
\end{array} }\,.
\end{equation}
Here, $\mathcal{I}$  and $\mathcal{J}$ are the charge and heat currents, respectively generated when an electrical 
bias $\Delta V=V_L-V_R$ and a thermal bias $\Delta T=T_L-T_R$ are applied to the left and right contacts. 
For definiteness, we take $V_{L/R} = V \pm \Delta V/2$ and $T_{L,R}=T\pm\Delta T/2$ 
where $V$ and $T$ are the bias and background temperature common to the two graphene flakes. 
In an explicit manner, the elements of the Onsager matrix elements are \cite{mahan1990many, PhysRevB.85.155443}
\begin{align}
\mathcal{L}_{11} &= \frac{e^2}{h}\int_{-\infty}^{+\infty} d\varepsilon~ 
                     \mathcal{T}(\varepsilon)
                     \left[-\frac{\partial{f(\varepsilon)}}{\partial{\varepsilon}} \right]_{T,\mu},  \\
\mathcal{L}_{12} &= \frac{e}{hT}\int_{-\infty}^{+\infty} d\varepsilon~
                     \mathcal{T}(\varepsilon) (\varepsilon - \mu)
                     \left[-\frac{\partial{f(\varepsilon)}}{\partial{\varepsilon}} \right]_{T,\mu},  \\
\mathcal{L}_{22} &= \frac{1}{hT}\int_{-\infty}^{+\infty} d\varepsilon~
                     \mathcal{T}(\varepsilon) (\varepsilon - \mu)^2
                     \left[-\frac{\partial{f(\varepsilon)}}{\partial{\varepsilon}} \right]_{T,\mu},
\end{align}
where $\mathcal{T}(\varepsilon)$ is the  transmission probability. 
The Onsager's coefficients ($\mathcal{L}_{ij}$)  are now connected to the linear electrical ($\mathcal{G}$), thermal ($\mathcal{K}$), and 
thermoelectrical ($\mathcal{L}_{12}$) conductances as
\begin{align}
\mathcal{G} &= \mathcal{L}_{11}, \label{eq:electrical_conductance} \, \\
\mathcal{K} &= \mathcal{L}_{22} - T\frac{\mathcal{L}_{12}^2}{\mathcal{L}_{11}}.
\end{align}
It is worthy to define the Seebeck coefficient or themopower in terms of the Onsager transport 
coefficients. The Seebeck coefficient reads
\begin{equation}
S = -\frac{\Delta V}{\Delta T}\Big|_{\mathcal{I}=0} = \frac{\mathcal{L}_{12}}{\mathcal{L}_{11}}\,,
\end{equation}
An important remark here is in order. Note that for sufficient low temperatures the thermal 
conductance is dominated by the electronic contribution and therefore we can neglect the phonon contribution. 
The efficiency of the heat-to-electricity (or vice-versa) conversion processes is measured by the figure of merit $ZT$
\begin{equation}
ZT = \frac{S^2 GT}{\mathcal{K}}.
\label{eq:ZT}
\end{equation}

All transport coefficients are expressed in terms of the transmission coefficient $\mathcal{T}(\varepsilon)$
\begin{align}
\mathcal{T}(\varepsilon) &= 2\pi \Gamma(\varepsilon) \sum_{\sigma}\rho_{d\sigma}(\varepsilon), 
\end{align}
where $\Gamma(\varepsilon)=\gamma_0 |\varepsilon|$ with $\gamma_0=\pi\tilde{V}^2\rho_0$ (here, $\rho_0 = 2/(\hbar v_F)^2$).
$\rho_{d\sigma}(\varepsilon)$ denotes the local DOS for the interacting level which can be found as 
\begin{equation}
\label{eq:LDOS}
\rho_{d\sigma}(\varepsilon) = -\frac{1}{\pi} \Im\left(G_{\sigma,\sigma}^{r}(\varepsilon)\right).
\end{equation}
\noindent  
$G_{\sigma,\sigma}^{r}$ is the retarded Green function for the interacting localized level. 
In the following, we derive a Green's function suitable for the Coulomb  blockade regime. 
To attain such a goal, we employ the EOM technique \cite{Lacroix98,Meir1993, Kashcheyevs2006} followed by a 
decoupling procedure (see below). We mention that 
Kondo correlation is not included in our decoupling scheme. However, our approach 
yields an excellent characterization of the transport properties of strongly interacting quantum dots for 
temperatures larger than the Kondo temperature or when the localized level is weakly coupled to the 
electrodes \cite{ 9780511470752}. Besides, it is known that for zero doped graphene Kondo effect is never built.

\subsection{Green's function}\label{sec:green_function}

A retarded Green's function $G^r(t)$ for fermionic operators $A$ and $B$ is defined as
\begin{equation}
\label{eq:green_definition}
G_{A,B}^{r}(t) \equiv \tgreen{A}{B} = -i \theta(t) \langle \{A(t),B(0)\}\rangle,
\end{equation}
whose EOM in energy space takes the following form
\begin{equation}
\label{eq:EOM}
\varepsilon^+ \wgreen{A}{B} + \wgreen{\left[\mathcal{H},A\right]}{B} = \langle\{A,B\}\rangle,
\end{equation}
with $\mathcal{H}$ being the Hamiltonian under consideration and $\varepsilon^+=\varepsilon+i0^+$.  
For the dot Green's function $G_{\sigma,\sigma}^{r}(\varepsilon) = \wgreen{d_{\sigma}}{d_{\sigma}^{\dagger}}$,
it is easy to show that
\begin{multline}
\label{eq:green-form-one}
\left( \varepsilon^+ - \varepsilon_{d\sigma} \right)G_{\sigma,\sigma}^{r}(\varepsilon)= 1 
+ U\wgreen{d_{\sigma}n_{\bar{\sigma}}}{d_{\sigma}^{\dag}} 
\\
+ \widetilde{V} \sum_{\alpha,s} \int_{-k_c}^{k_c} dk~\sqrt{|k|}\wgreen{c_{\alpha sk\sigma}}{d_{\sigma}^{\dag}} \,.
\end{multline}
The equation for $\wgreen{c_{\alpha sk\sigma}}{d_{\sigma}^{\dag}}$ is found to be
\begin{equation}\label{eq:F-function-one}
\wgreen{c_{\alpha sk\sigma}}{d_{\sigma}^{\dag}} 
= \frac{\widetilde{V}\sqrt{|k|}}{\varepsilon^+ - \varepsilon_k}G_{\sigma,\sigma}^{r}(\varepsilon)
\end{equation}
such that the dot Green's function becomes
\begin{equation}\label{eq:green-noapprox}
\left(\varepsilon - \varepsilon_{d\sigma} - \Sigma_0^r(\varepsilon) \right)G_{\sigma,\sigma}^{r}(\varepsilon)
= 1 + U\wgreen{d_{\sigma}n_{\bar{\sigma}}}{d_{\sigma}^{\dag}},
\end{equation}
where $\Sigma_0^r(\varepsilon)= \widetilde{V}^2 \sum_{\alpha,s} \int_{-k_c}^{k_c} dk\,\frac{|k|}{\varepsilon^{+}-\varepsilon_{k}} $ 
is the self-energy due to the hybridization between graphene contacts and localized level. 
The self-energy is evaluated as
\begin{equation}\label{eq:auto-energy}
\Sigma_0^r(\varepsilon) = -\eta \left[\varepsilon \ln\left| \frac{D^2 - \varepsilon^2}{\varepsilon^2} \right| + i\pi|\varepsilon|\Theta(D - |\varepsilon|)\right]
\end{equation}
where  $\eta = 2( \tilde{V}/\hbar v_F )^2 = \gamma_0/\pi$. 

In order to get the Coulomb blockade solution, we need to calculate the EOM for $\wgreen{d_{\sigma}n_{\bar{\sigma}}}{d_{\sigma}^{\dag}}$.
It is given by
\begin{widetext}
\begin{multline}\label{eq:GammaEOM-2}
 (\varepsilon^{+}-\varepsilon_{d\sigma}-U)\wgreen{d_{\sigma}n_{\bar{\sigma}}}{d_{\sigma}^{\dag}} 
= \langle n_{\bar{\sigma}}\rangle
+ \widetilde{V}\sum_{\alpha,s} \int_{-k_c}^{k_c}dk~\sqrt{|k|}
\left[\wgreen{c_{\alpha sk\sigma}n_{\bar{\sigma}}}{d_{\sigma}^{\dagger}} 
+ \wgreen{d_{\bar{\sigma}}^{\dagger}c_{\alpha sk\bar{\sigma}}d_{\sigma}}{d_{\sigma}^{\dagger}}
\right. \\ \left.
- \wgreen{c_{\alpha sk\bar{\sigma}}^{\dagger}d_{\bar{\sigma}}d_{\sigma}}{d_{\sigma}^{\dagger}}\right] \,.
\end{multline}
\end{widetext}
We keep only the correlation $\wgreen{c_{\alpha sk\sigma}n_{\bar{\sigma}}}{d_{\sigma}^{\dagger}}$ on the right hand side 
and calculate its EOM which can be approximated as
\begin{equation}\label{eq:X-EOM-CB}
(\varepsilon^{+}-\varepsilon_k)\wgreen{c_{\alpha sk\sigma}n_{\bar{\sigma}}}{d_{\sigma}^{\dagger}}  
\approx \widetilde{V} \sqrt{|k|}\wgreen{d_{\sigma}n_{\bar{\sigma}}}{d_{\sigma}^{\dag}} \,.
\end{equation}
Gathering all the results, we obtain
\begin{equation}\label{eq:greenFunction-CB}
G_{\sigma,\sigma}^r(\varepsilon) 
= \left[\frac{1-\langle n_{\bar{\sigma}}\rangle}{\varepsilon-\varepsilon_{d\sigma}-\Sigma_0^r(\varepsilon)}
+ \frac{\langle n_{\bar{\sigma}}\rangle}{\varepsilon-\varepsilon_{d\sigma}-U-\Sigma_0^r(\varepsilon)}\right] \,.
\end{equation}
The poles of $G_{\sigma,\sigma}^r(\varepsilon)$ are located around $\varepsilon_{d\sigma}$ and $\varepsilon_{d\sigma}+U$
such that this solution properly describes the Coulomb blockade regime.
The dot occupation must be calculated self-consistently using
\begin{equation}
\label{eq:occupation}
\langle n_{\sigma}\rangle = \int_{-D}^{+D} d\varepsilon~ f(\varepsilon)\left[-\frac{1}{\pi}\Im \left(G_{\sigma,\sigma}^r(\varepsilon)\right)\right].
\end{equation}
where $f(\varepsilon) = 1/(\exp(\varepsilon/k_BT)+1)$.
In the absence of a magnetic field, 
we expect the paramagnetic solution $\langle n_\sigma\rangle = \langle n_{\bar\sigma}\rangle$. 

In the next section, we present our results for the transport coefficients as a function of  the dot level (\ref{sec:transportvsEd}) and 
the background temperature (\ref{sec:transportvskBT}). 
The Seebeck coefficient and the ZT figure of merit are also analyzed.
Finally, we illustrate that Dirac-like setup exhibits strong departures from the Wiedemann-Franz law. 

\section{Results}
\begin{figure}
\centering
\includegraphics[width=.5\textwidth]{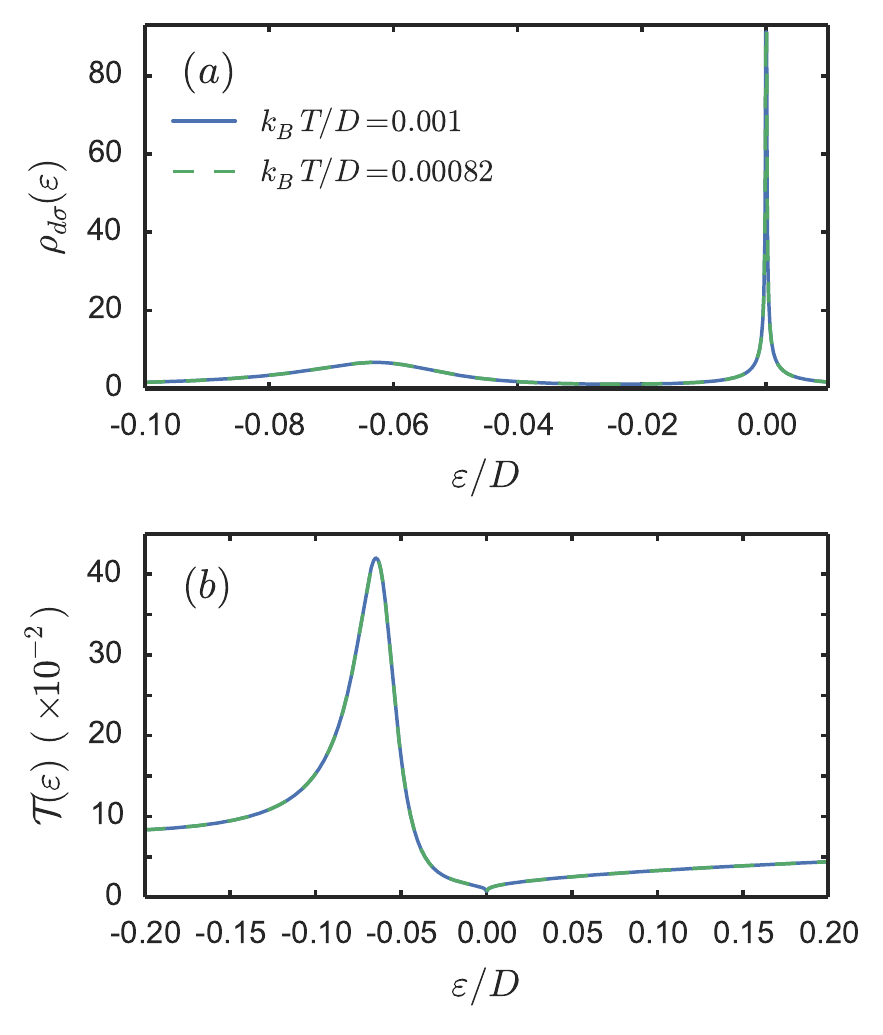}%
\caption{(a) Dot density of states $\rho_{d\sigma}(\varepsilon)$ and (b) transmission coefficient $\mathcal{T}(\varepsilon)$. 
We consider two distinct temperatures: $k_BT/D=0.001$ (solid line) and $8.2\times 10^{-5}$ (dotted line). 
Rest of parameters: $\varepsilon_{d\sigma} = -U$, $\eta = 0.1$, $D\sim 7\,eV$, and $U/D = 0.1$.}
\label{fig:LDOSTrvsEd_T}
\end{figure}

Our results for the transport properties of a graphene based quantum dot are shown.  
Unless it is indicated, we consider the strong Coulomb blockade regime ($U=0.1D$) and investigate how the transport properties 
depend on when the dot level is varied and when the background temperature is tuned. 
In these two cases, we explore the electrical and thermal conductances, and the Seebeck and ZT coefficients. \label{sec:results}
In order to better understand the transport properties of our system, in Fig.~\ref{fig:LDOSTrvsEd_T} we briefly discuss the 
behavior of the dot DOS and the transmission coefficient. 
It is shown that for $\varepsilon_{d\sigma}=-U$ the DOS displays two resonances located at 
$\varepsilon\approx \varepsilon_{d\sigma}+\Re  e\Sigma_0(\varepsilon)$ and $\varepsilon\approx \varepsilon_{d\sigma}+U+\Re e\Sigma_0(\varepsilon)$. 
The resonance around the Fermi energy ($\varepsilon_F=0$) is very narrow in comparison with the high 
energy resonance which is more wider. This is explained by the fact that the hybridization depends 
strongly on energy.  In addition, the dot DOS at the Fermi energy becomes maximum. Nevertheless, even if the local 
DOS reaches the highest value at $\varepsilon_F$, the transmission coefficient, evaluated at the Fermi energy, 
vanishes. This is illustrated in Fig. \ref{fig:LDOSTrvsEd_T}(b). The transmission coefficient is the product 
of the contact and dot density of states. The former vanishes exactly at the Fermi energy giving rise to a null 
transmission coefficient even though the dot DOS attains the highest value. The fact that the transmission coefficient 
gets a zero value at $\varepsilon_F$ yields a two-dip structure for the electrical conductance versus the 
dot level position as shown below. 

\subsection{Transport coefficients versus the gate voltage}\label{sec:transportvsEd}
\begin{figure}
\centering
\includegraphics[width=.5\textwidth]{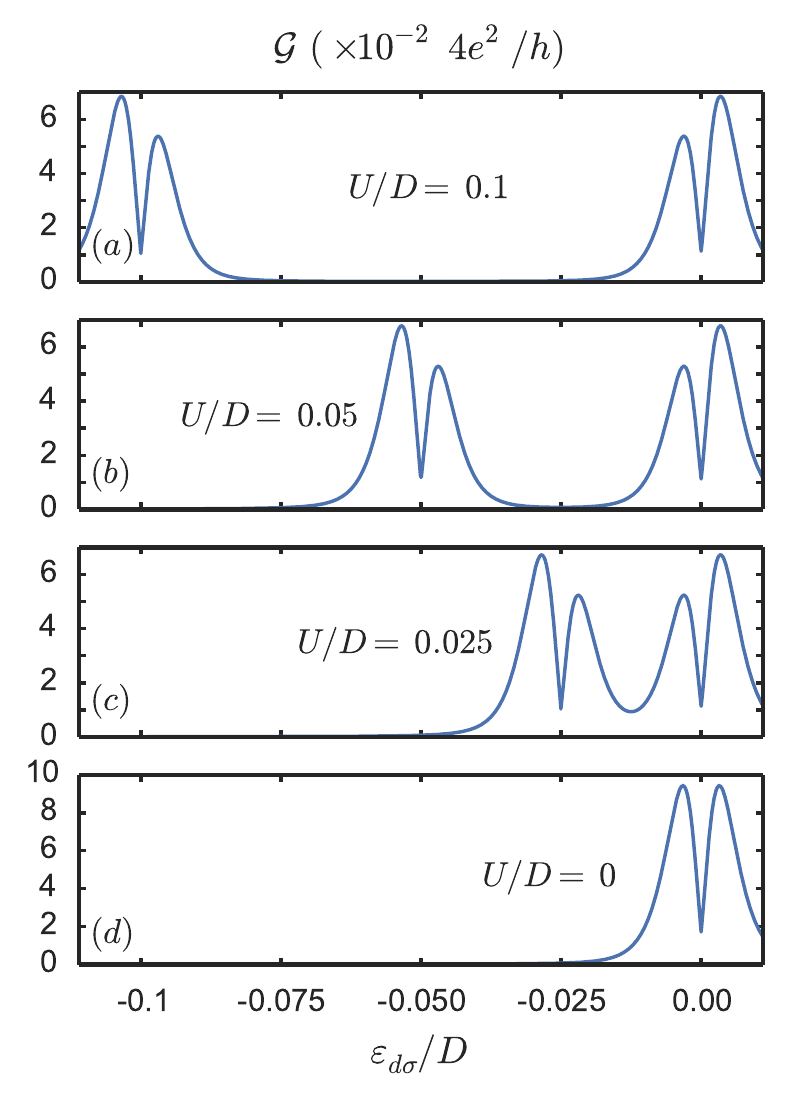}
\caption{Electric conductance $\mathcal{G}$ as a function of $\varepsilon_{d\sigma}$ for $U/D = 0.1$ (a), $0.05$ (b), $0.025$ (c), and $0$ (d). 
Rest of parameters: $\eta = 0.1$, $D\sim 7\,eV$,  and $k_BT/D= 0.001$.}
\label{fig:GvsEd_U}
\end{figure}

Now we discuss the electrical and thermal conductances when the dot level is tuned.
We start by showing the evolution of the electrical conductance $\mathcal{G}$ as a function of the dot 
level for several values of the on-site interaction $U$ (see Fig.~\ref{fig:GvsEd_U}). 
For the noninteracting case ($U/D=0$), the conductance exhibits a dip (antiresonance) at the contact Fermi energy $\varepsilon_F$.
In general, the resonance happens when the localized level alignes with $\varepsilon_F$. 
Since the contact DOS vanishes at $\varepsilon=\varepsilon_F$, we observe the antiresonance.  
In the presence of $U$, there are two effective levels located at $\varepsilon_d$ and $\varepsilon_d+U$.
When these two levels align with $\varepsilon_F$, the transmission again vanishes due to the nonavailability of states of the  graphene contacts at $\varepsilon_F$. 
The transmission shows a dip that resembles a Fano singularity originated when an interference event occurs. 
However, notice that the origin of the dip in our setup is due to the lack of electronic states of the graphene contacts 
instead of being produced by an interference between different paths. 
Fano resonances are usually found in double dot systems \cite{Fano1}. 
It is remarkable that elevated values for the Seebeck coefficient are encountered in double dots where Fano antiresonances take place \cite{Fano2}. Here, as shown later, we find similar behavior, i.e., the occurrence of dips at the transmission leads to high thermopower values.  

Another aspect of $\mathcal{G}$ is that the peaks near to the dips are not symmetric.
Around $\varepsilon_{d\sigma} = -U$, that is, the two maxima are different in their heights. 
This results from the fact that the localized level DOS is not symmetric with respect to the resonance points. 
Notice that this does not happens in normal contacts. 

\begin{figure}
\centering
\includegraphics[width=.5\textwidth]{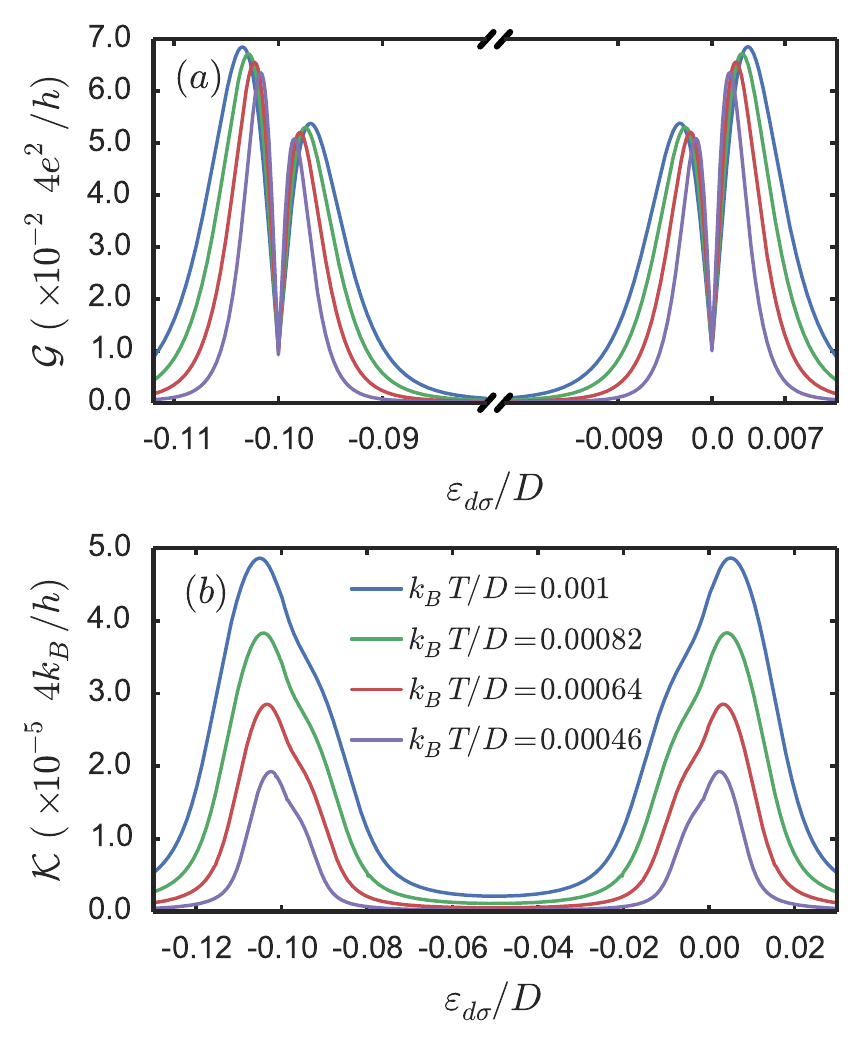}
\caption{(a) Electric conductance $\mathcal{G}$ and (b) thermal conductance $\mathcal{K}$ versus $\varepsilon_{d\sigma}$ for various $k_BT/D$. Rest of parameters: $\eta = 0.1$, $D\sim 7\,eV$,  and $U/D = 0.1$.}
\label{fig:GKvsEd_T}
\end{figure}

\begin{figure}
\centering
\includegraphics[width=.5\textwidth]{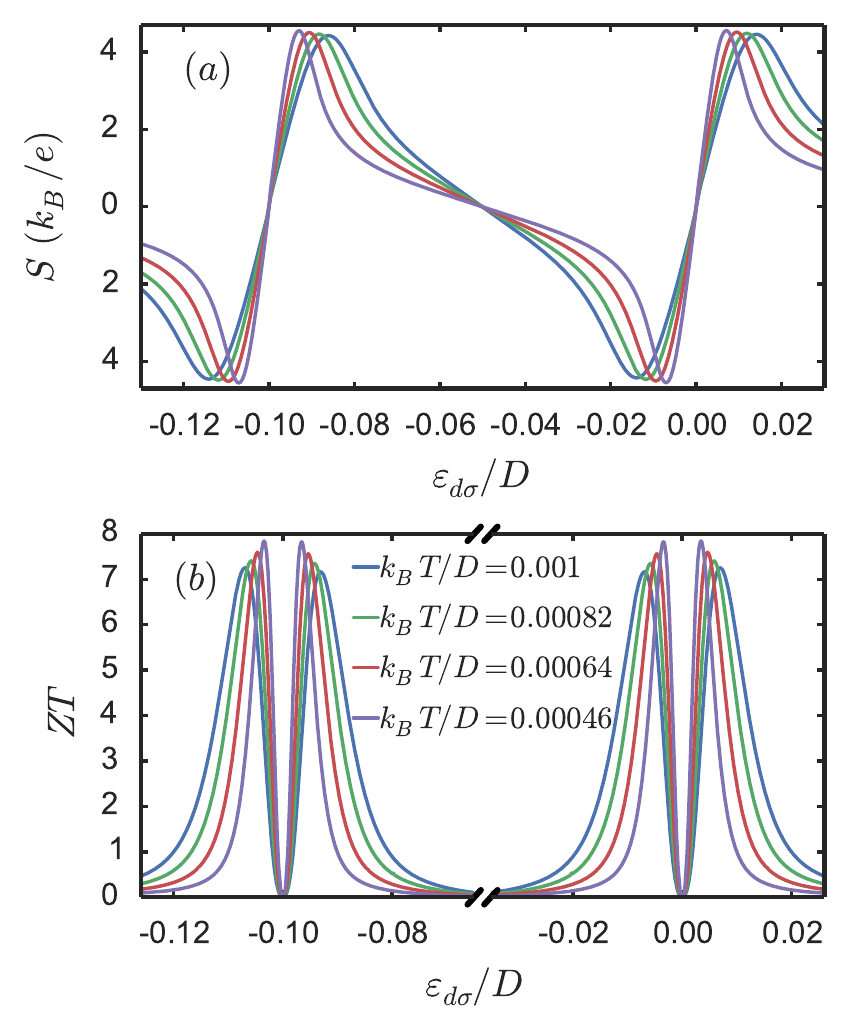}
\caption{(a) Seebeck coefficient $S$ and (b) figure of merit $ZT$ versus $\varepsilon_{d\sigma}$ for various $k_BT/D$. Rest of parameters: $\eta = 0.1$, $D\sim 7\,eV$,  and $U/D = 0.1$.}
\label{fig:SZTvsEd_T}
\end{figure}

We focus on the strong interacting case where Coulomb blockade phenomenon is better exhibited. We choose $U/D=0.1$. 
Our purpose is to analyze the temperature dependence of the electrical $\mathcal{G}$ and thermal $\mathcal{K}$ conductances as shown 
in Fig.~\ref{fig:GKvsEd_T}(a) and (b), respectively.  
It is observed that $\mathcal{G}$ has moderate temperature dependences as a function of the gate voltage. 
In the middle of the valley, the thermal activation is not sufficient to draw a quantitative change in $\mathcal{G}$.  
On the contrary, $\mathcal{K}$ shows more dramatic behaviors as a function of $\varepsilon_{d\sigma}$.  The first 
remarkable fact is the absence of the dips at the resonance points. Even though electrical transport is blockaded at 
the resonance points, the heat transports do not follow such behaviors.  

\begin{figure}
\centering
\includegraphics[width=.5\textwidth]{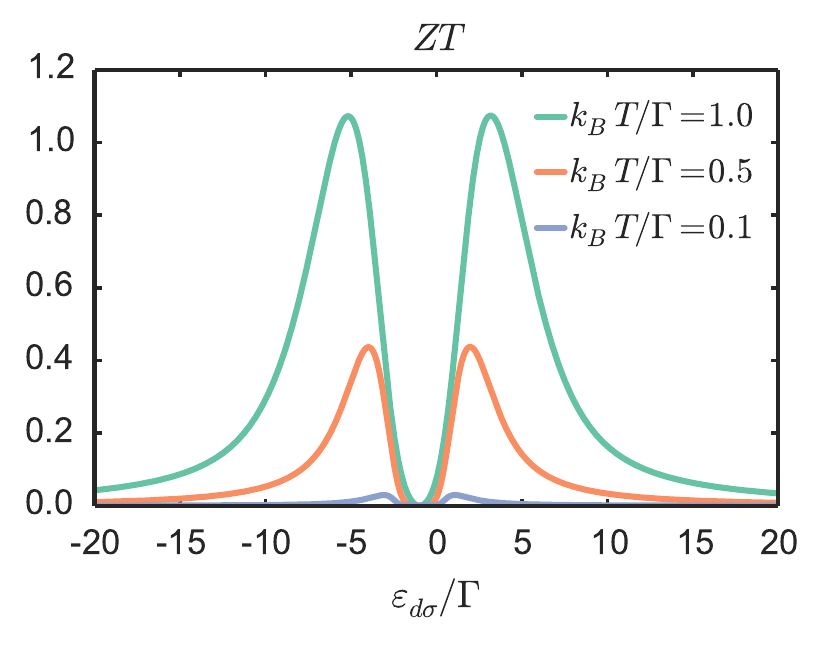}%
\caption{Figure of merit $ZT$ as a function of $\varepsilon_{d\sigma}$ for various $k_BT/D$. 
Metallic contacts are considered instead of graphene contacts.
Rest of parameters: $\Gamma = 1$, $D/\Gamma = 100$, and $U/\Gamma = 2$.}
\label{fig:ZTvsEdmetal}
\end{figure}

Importantly, $\mathcal{K}$ is three orders of magnitude smaller than $\mathcal{G}$.  
This is a notable feature when compared with normal conductors that possess high electrical and thermal conductances at the same time. 
Good thermoelectrical devices are those that display poor thermal conductances and high electrical conductances just like 
our graphene transistor. 
In metallic conductors, the Wiedemann-Franz law 
\begin{equation}
\frac{\mathcal{K}}{\mathcal{G}T} = \frac{\pi^2}{3}\left(\frac{k_B}{e}\right)^2
\end{equation}
is satisfied.
By utilizing graphene contacts, such a relation between $\mathcal{G}$ and $\mathcal{K}$ does not hold anymore. 
Consequently, the Seebeck coefficient attains much higher values. 
This fact is precisely illustrated in Fig.~\ref{fig:SZTvsEd_T}(a) where the Seebeck coefficient is displayed as a function of 
$\varepsilon_{d\sigma}$ for various background temperatures $T$. 
It is shown that the Seebeck coefficient reaches quite high values $S\gtrsim 4$. 
As expected, $S$ vanishes at the resonance points $\varepsilon_{d\sigma}=\varepsilon_F$ and 
$\varepsilon_{d\sigma}+U=\varepsilon_F$. Around these points, $S$ is an odd function of $\varepsilon_{d\sigma}$. 
Related with $S$ is the figure of merit $ZT$. 
In Eq.~\eqref{eq:ZT}, the thermal conductance $\mathcal{K}$ contains the electronic and phonon contributions (i.e., $\mathcal{K} = \mathcal{K}_e + \mathcal{K}_{ph}$). 
Through this work, $\mathcal{K}_{ph}$ is neglected since we assume the low temperature limit ($T< 80 K$). 
Besides, in graphene,  the thermal conductance due to phonons can be tailored to get very low values even at room temperature by including anti-dots or nanoribbons in it \cite{kphonon}. Therefore, our calculations do not consider the phonon contribution to the thermal conductance. 
The reported large values of $ZT$ indicate that heat-to-electricity conversion process is performed with high efficiency. 
Good thermoelectrical conductors can exhibit $ZT\gtrsim1$. Our device shows $ZT$ values close to $8$ as illustrated in Fig. \ref{fig:SZTvsEd_T}(b) which 
is a remarkable fact for practical applications. 
These results for the figure of merit $ZT$ can be compared with those achieved for the case of a localized level tunnel coupled to normal contacts with  an energy independent tunneling rate $\Gamma$.  
Such comparison  is performed in Fig. \ref{fig:ZTvsEdmetal} where the ZT values for the normal dot case are displayed. 
We observe that much smaller $ZT$ values are reached.

\begin{figure}
\centering
\includegraphics[width=.5\textwidth]{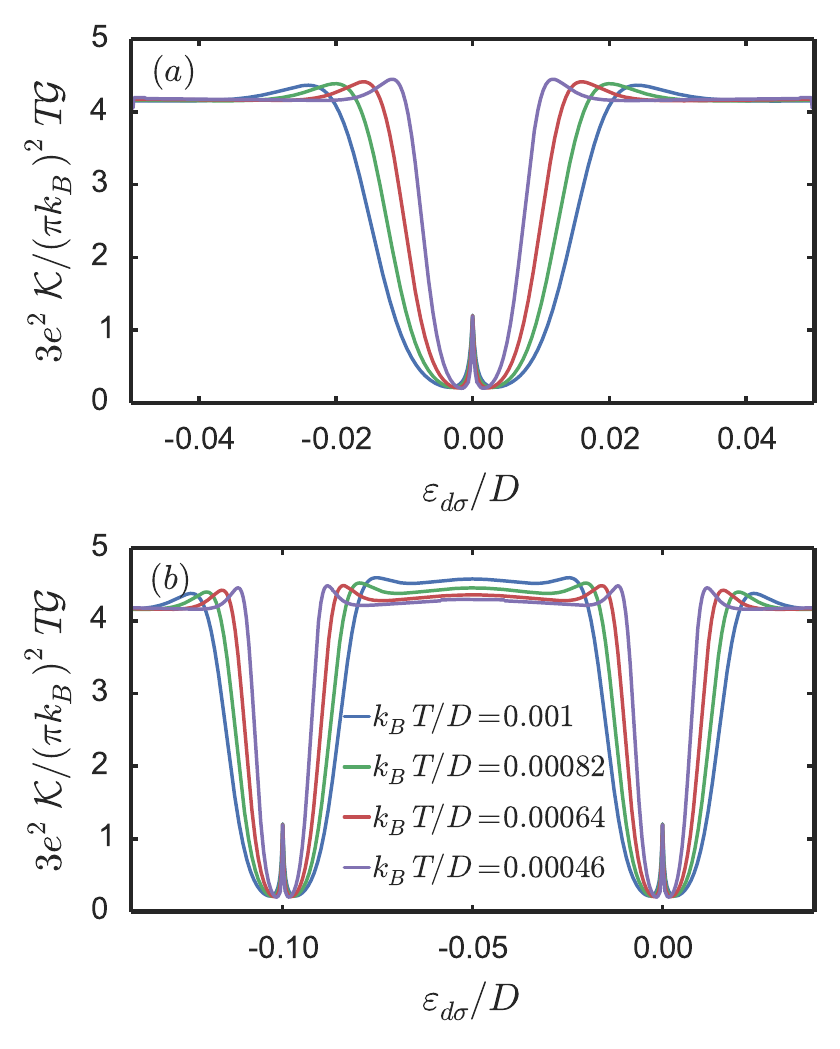}%
\caption{Violation of the Wiedemann-Franz law as a function of $\varepsilon_{d\sigma}$ for $U/D = 0$ (a) and $0.1$ (b). 
As shown, several temperatures are considered.  Rest of parameters: $\eta = 0.1$, $D\sim 7\,eV$.}
\label{fig:WFvsEd_T}
\end{figure}

In Fig.~\ref{fig:WFvsEd_T}, we investigate the Wiedemann-Franz law.
As anticipated, the  Wiedemann-Franz law is not fulfilled. 
To pinpoint the main source of such violation, we consider $U=0$ and $U\ne 0$ cases. 
It is observed that the violation of the Wiedemann-Franz is nothing to do with $U$. 
We thus conclude that the departure from the Wiedemann-Franz law is due to the Dirac-like energy dispersion relation of the contacts.

\subsection{Transport properties versus background temperature}\label{sec:transportvskBT}

\begin{figure}
\centering
\includegraphics[width=.5\textwidth]{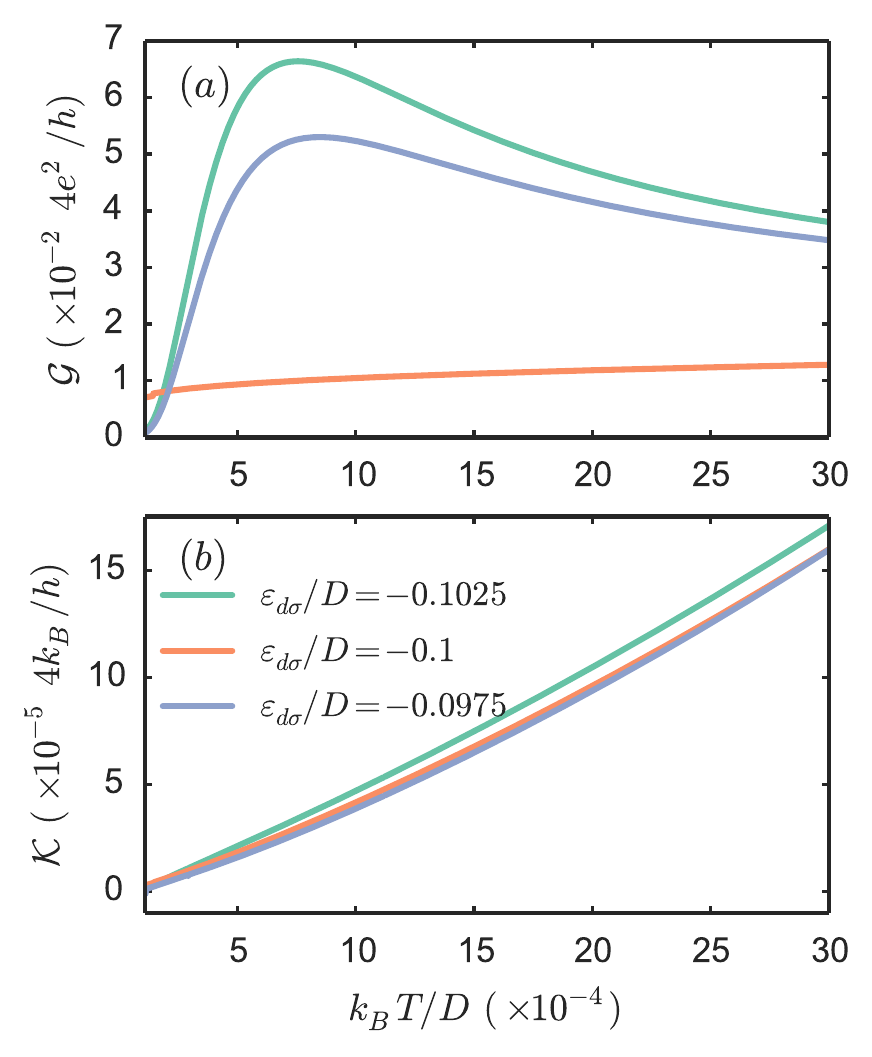}
\caption{(a) Electrical conductance $\mathcal{G}$ and (b) thermal conductance $\mathcal{K}$ versus the background temperature $k_BT$ for different  values of $\varepsilon_{d\sigma}$.  
At $\varepsilon_{d\sigma}/D = -0.1025$ and $-0.0975$, maximum values are reached in $\mathcal{G}$,
while for $\varepsilon_{d\sigma}/D =-0.1$ the dip is observed in  $\mathcal{G}$ (see Fig. \ref{fig:GvsEd_U}(a)). 
Rest of parameters: $\eta = 0.1$, $D\sim 7\,eV$,  and $U/D= 0.1$.}
\label{fig:GKvsT_Ed}
\end{figure}

\begin{figure}
\centering
\includegraphics[width=.5\textwidth]{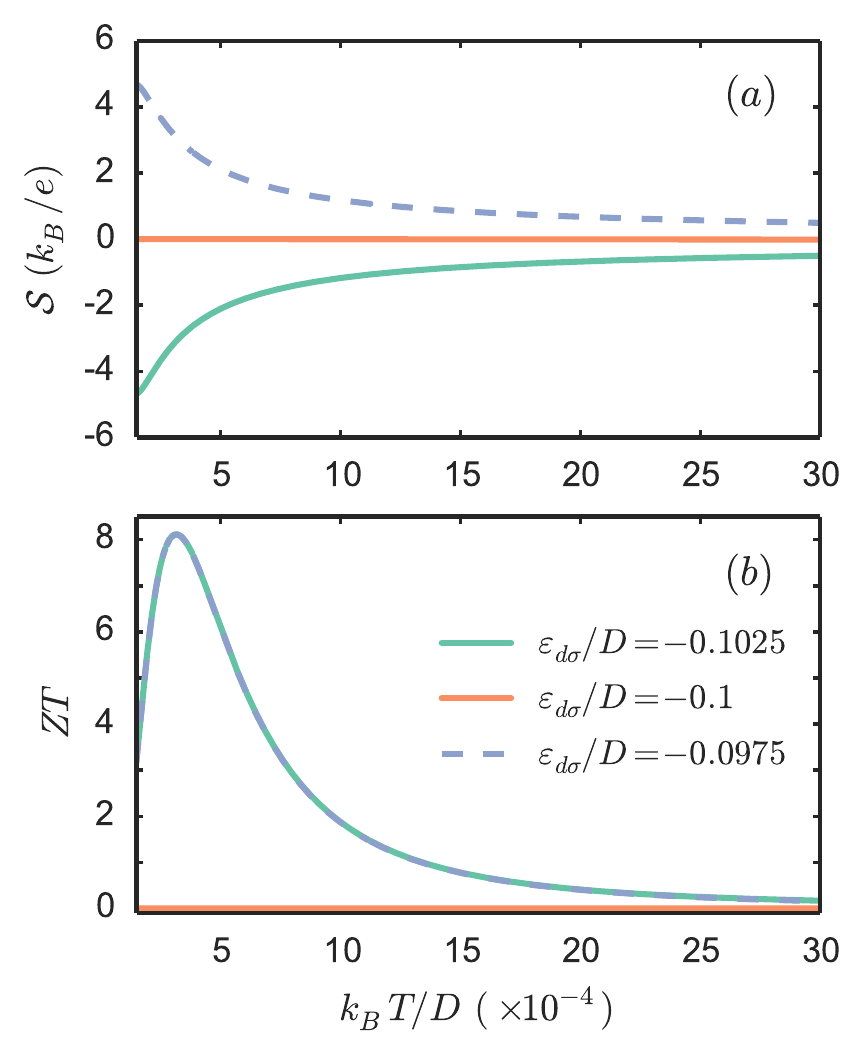}
\caption{(a) Seebeck coefficient $S$ and (b) figure of merit $ZT$ as a function of $k_BT$ when $\varepsilon_{d\sigma}/D$ is changed. 
Rest of parameters: $\eta = 0.1$, $D\sim 7\,eV$,  and $U/D = 0.1$.}
\label{fig:SZTvsT_Ed}
\end{figure}

We now examine $\mathcal{G}$, $\mathcal{K}$, $S$, and $ZT$ as a function of the background temperature $T$. 
Figure~\ref{fig:GKvsT_Ed}(a) displays the electrical conductance $\mathcal{G}$ when $k_BT$ is varied for three dot level 
positions ($\varepsilon_{d\sigma}/D=-0.1025$, $-0.1$, and $-0.0975$). 
At the resonance point ($\varepsilon_{d\sigma}/D=-0.1$), $\mathcal{G}(T)$ is almost flat getting a very 
small value due to the lack of electronic states at the contacts for energies close to the Dirac point. 
For the other dot level positions, $\mathcal{G}(T)$ shows a maximum with its 
position that slightly depends on the value of $\varepsilon_{d\sigma}$. The maximum occurs when 
$k_BT\approx \delta$ which denotes the peak width of the dot DOS for each level position. We recall that $\delta$
is a function of $\varepsilon_{d\sigma}$. On the other hand, $\mathcal{K}(T)$ increases linearly with $T$ irrespective of $\varepsilon_{d\sigma}$. 

We end up by exploring the behavior of the Seebeck and $ZT$ coefficients. The Seebeck coefficient [see 
Fig.~\ref{fig:SZTvsT_Ed}(a)]  vanishes for the resonance point as expected due to the particle-hole symmetry without respect to temperature. For 
the other two localized level values the Seebeck coefficient takes opposite signs. This is due to the fact that  
for $\varepsilon_{d\sigma}/D=-0.1025$ the DOS has more weight at the left of the Fermi energy (so-called hole-
like transport) whereas for $\varepsilon_{d\sigma}/D=-0.0975$ the situation is reversed (electron-like 
transport). Here is remarkable the fact that at low and moderate temperatures $S$ reaches considerable high values 
(above $S>1$). Finally the $ZT$ coefficient is proportional to $S^2$ and therefore is not sign sensitive. For 
$\varepsilon_{d\sigma}/D=-0.1$ the $ZT$ is always null whereas for  the other two level positions $ZT$ shows 
a similar behavior. The $ZT$ has an optimized value at certain temperature scale $k_B T^*$ 
reaching the value close to $ZT\approx 8$.  

\section{Conclusions}
In closing, we have analyzed the linear electrical, thermal, and thermoelectrical transports for an interacting localized 
level coupled to Dirac-like electrodes made by graphene. Our results support the fact that Dirac-like electrodes lead 
to much better thermolectrical devices with very high values for the Seebeck and $ZT$ coefficients. The latter can reach values of $ZT\approx 8$.  Indeed, due to the 
strong energy dependence of the density of states of the electrodes our graphene device display a rather low thermal 
conductance in contrast with the high and moderate values reached for the electrical one. 
These two facts yield high thermoelectrical efficiencies. 
Besides, such energy dependence is responsible for the violation of the Wiedemann-Franz law. 
In addition, we report results for the conductances as a function of the background temperature. 
Remarkably, $\mathcal{G}(T)$ and $ZT(T)$ shows an optimal value which depends on the dot level position. 
On the contrary, $\mathcal{K}(T)$ grows monotonically with temperature. 
Such growing behavior depends on the dot level position. 
Finally, as expected the Seebeck coefficient reverses its sign when transport occurs via quasi-holes and changes to quasi-electrons (from $\varepsilon_{d\sigma}+U<0$ to $\varepsilon_{d\sigma}+U>0$).  We believe that the reported values for the $ZT$ in graphene based quantum dots will motivate its experimental verification using the present technology.
 
\section{Acknowledgement}
Work supported by MINECO Grant No. FIS2014-52564  and  by the
Kavli Institute for Theoretical Physics through NSF grant PHY11-25915.



\begin{thebibliography}{44}
\expandafter\ifx\csname natexlab\endcsname\relax\def\natexlab#1{#1}\fi
\expandafter\ifx\csname bibnamefont\endcsname\relax
  \def\bibnamefont#1{#1}\fi
\expandafter\ifx\csname bibfnamefont\endcsname\relax
  \def\bibfnamefont#1{#1}\fi
\expandafter\ifx\csname citenamefont\endcsname\relax
  \def\citenamefont#1{#1}\fi
\expandafter\ifx\csname url\endcsname\relax
  \def\url#1{\texttt{#1}}\fi
\expandafter\ifx\csname urlprefix\endcsname\relax\def\urlprefix{URL }\fi
\providecommand{\bibinfo}[2]{#2}
\providecommand{\eprint}[2][]{\url{#2}}

\bibitem[{\citenamefont{Neto et~al.}(2009{\natexlab{a}})\citenamefont{Neto,
  Guinea, Peres, Novoselov, and Geim}}]{RevModPhys.81.109}
\bibinfo{author}{\bibfnamefont{A.~H.~C.} \bibnamefont{Neto}},
  \bibinfo{author}{\bibfnamefont{F.}~\bibnamefont{Guinea}},
  \bibinfo{author}{\bibfnamefont{N.~M.~R.} \bibnamefont{Peres}},
  \bibinfo{author}{\bibfnamefont{K.~S.} \bibnamefont{Novoselov}},
  \bibnamefont{and} \bibinfo{author}{\bibfnamefont{A.~K.} \bibnamefont{Geim}},
  \bibinfo{journal}{Rev. Mod. Phys.} \textbf{\bibinfo{volume}{81}}
  (\bibinfo{year}{2009}{\natexlab{a}}).

\bibitem[{\citenamefont{Geim and Novoselov}(2007)}]{Geim_2007}
\bibinfo{author}{\bibfnamefont{A.~K.} \bibnamefont{Geim}} \bibnamefont{and}
  \bibinfo{author}{\bibfnamefont{K.~S.} \bibnamefont{Novoselov}},
  \bibinfo{journal}{Nature Materials} \textbf{\bibinfo{volume}{6}},
  \bibinfo{pages}{183} (\bibinfo{year}{2007}).

\bibitem[{\citenamefont{Wallace}(1947)}]{Wallace47}
\bibinfo{author}{\bibfnamefont{P.~R.} \bibnamefont{Wallace}},
  \bibinfo{journal}{Phys. Rev.} \textbf{\bibinfo{volume}{71}},
  \bibinfo{pages}{622} (\bibinfo{year}{1947}).

\bibitem[{\citenamefont{Lomer}(1955)}]{Lomer1955}
\bibinfo{author}{\bibfnamefont{W.~M.} \bibnamefont{Lomer}},
  \bibinfo{journal}{Proceedings of the Royal Society of London A: Mathematical,
  Physical and Engineering Sciences} \textbf{\bibinfo{volume}{227}},
  \bibinfo{pages}{330} (\bibinfo{year}{1955}).

\bibitem[{\citenamefont{McClure}(1956)}]{McClure1956}
\bibinfo{author}{\bibfnamefont{J.~W.} \bibnamefont{McClure}},
  \bibinfo{journal}{Phys. Rev.} \textbf{\bibinfo{volume}{104}},
  \bibinfo{pages}{666} (\bibinfo{year}{1956}).

\bibitem[{\citenamefont{McClure}(1957)}]{McClure1957}
\bibinfo{author}{\bibfnamefont{J.~W.} \bibnamefont{McClure}},
  \bibinfo{journal}{Phys. Rev.} \textbf{\bibinfo{volume}{108}},
  \bibinfo{pages}{612} (\bibinfo{year}{1957}).

\bibitem[{\citenamefont{Slonczewski and Weiss}(1958)}]{Slonczewski1958}
\bibinfo{author}{\bibfnamefont{J.~C.} \bibnamefont{Slonczewski}}
  \bibnamefont{and} \bibinfo{author}{\bibfnamefont{P.~R.} \bibnamefont{Weiss}},
  \bibinfo{journal}{Phys. Rev.} \textbf{\bibinfo{volume}{109}},
  \bibinfo{pages}{272} (\bibinfo{year}{1958}).

\bibitem[{\citenamefont{F. and Parravicini}(1967)}]{Bassani1967}
\bibinfo{author}{\bibfnamefont{F.~B.} \bibnamefont{F.}} \bibnamefont{and}
  \bibinfo{author}{\bibfnamefont{G.~P.} \bibnamefont{Parravicini}},
  \bibinfo{journal}{Nuovo Cimento B Serie} \textbf{\bibinfo{volume}{50}},
  \bibinfo{pages}{95} (\bibinfo{year}{1967}).

\bibitem[{\citenamefont{DiVincenzo and Mele}(1984)}]{DiVincenzo1984}
\bibinfo{author}{\bibfnamefont{D.}~\bibnamefont{DiVincenzo}} \bibnamefont{and}
  \bibinfo{author}{\bibfnamefont{E.~J.} \bibnamefont{Mele}},
  \bibinfo{journal}{Phys. Rev. B} \textbf{\bibinfo{volume}{29}}
  (\bibinfo{year}{1984}).

\bibitem[{\citenamefont{Nebel}(2013)}]{Nebel_2013}
\bibinfo{author}{\bibfnamefont{C.~E.} \bibnamefont{Nebel}},
  \bibinfo{journal}{Nature Materials} \textbf{\bibinfo{volume}{12}},
  \bibinfo{pages}{690} (\bibinfo{year}{2013}).

\bibitem[{\citenamefont{Lundeberg and Folk}(2014)}]{Lundeberg_2014}
\bibinfo{author}{\bibfnamefont{M.~B.} \bibnamefont{Lundeberg}}
  \bibnamefont{and} \bibinfo{author}{\bibfnamefont{J.~A.} \bibnamefont{Folk}},
  \bibinfo{journal}{Science} \textbf{\bibinfo{volume}{346}},
  \bibinfo{pages}{422} (\bibinfo{year}{2014}).

\bibitem[{\citenamefont{Stampfer et~al.}(2008)\citenamefont{Stampfer,
  G\"{u}ttinger, Molitor, Graf, Ihn, and Ensslin}}]{Stampfer_2008}
\bibinfo{author}{\bibfnamefont{C.}~\bibnamefont{Stampfer}},
  \bibinfo{author}{\bibfnamefont{J.}~\bibnamefont{G\"{u}ttinger}},
  \bibinfo{author}{\bibfnamefont{F.}~\bibnamefont{Molitor}},
  \bibinfo{author}{\bibfnamefont{D.}~\bibnamefont{Graf}},
  \bibinfo{author}{\bibfnamefont{T.}~\bibnamefont{Ihn}}, \bibnamefont{and}
  \bibinfo{author}{\bibfnamefont{K.}~\bibnamefont{Ensslin}},
  \bibinfo{journal}{App. Phys. Lett.} \textbf{\bibinfo{volume}{92}},
  \bibinfo{pages}{012102} (\bibinfo{year}{2008}).

\bibitem[{\citenamefont{Tan et~al.}(2011)\citenamefont{Tan, Liu, Lu, and
  Yang}}]{Tan_2011}
\bibinfo{author}{\bibfnamefont{Z.}~\bibnamefont{Tan}},
  \bibinfo{author}{\bibfnamefont{G.}~\bibnamefont{Liu}},
  \bibinfo{author}{\bibfnamefont{L.}~\bibnamefont{Lu}}, \bibnamefont{and}
  \bibinfo{author}{\bibfnamefont{C.}~\bibnamefont{Yang}},
  \bibinfo{journal}{Sci. China Phys. Mech. Astron.}
  \textbf{\bibinfo{volume}{55}}, \bibinfo{pages}{7} (\bibinfo{year}{2011}).

\bibitem[{\citenamefont{Sols et~al.}(2007)\citenamefont{Sols, Guinea, and
  Neto}}]{Sols_2007}
\bibinfo{author}{\bibfnamefont{F.}~\bibnamefont{Sols}},
  \bibinfo{author}{\bibfnamefont{F.}~\bibnamefont{Guinea}}, \bibnamefont{and}
  \bibinfo{author}{\bibfnamefont{A.~H.~C.} \bibnamefont{Neto}},
  \bibinfo{journal}{Phys. Rev. Lett.} \textbf{\bibinfo{volume}{99}}
  (\bibinfo{year}{2007}).

\bibitem[{\citenamefont{Murali}(2012)}]{Murali_2012}
\bibinfo{author}{\bibfnamefont{R.}~\bibnamefont{Murali}}, in
  \emph{\bibinfo{booktitle}{Graphene Nanoelectronics: From Materials to
  Circuits}} (\bibinfo{publisher}{Springer-Verlag New York},
  \bibinfo{year}{2012}).

\bibitem[{\citenamefont{Hewson}(1993)}]{9780511470752}
\bibinfo{author}{\bibfnamefont{A.~C.} \bibnamefont{Hewson}},
  \emph{\bibinfo{title}{The Kondo problem to Heavy Fermions}}
  (\bibinfo{publisher}{Cambridge University Press}, \bibinfo{year}{1993}).

\bibitem[{\citenamefont{Cronenwett et~al.}(1998)\citenamefont{Cronenwett,
  Oosterkamp, and Kouwenhoven}}]{Cronenwett24071998}
\bibinfo{author}{\bibfnamefont{S.~M.} \bibnamefont{Cronenwett}},
  \bibinfo{author}{\bibfnamefont{T.~H.} \bibnamefont{Oosterkamp}},
  \bibnamefont{and} \bibinfo{author}{\bibfnamefont{L.~P.}
  \bibnamefont{Kouwenhoven}}, \bibinfo{journal}{Science}
  \textbf{\bibinfo{volume}{281}}, \bibinfo{pages}{540} (\bibinfo{year}{1998}).

\bibitem[{\citenamefont{Goldhaber-Gordon
  et~al.}(1998)\citenamefont{Goldhaber-Gordon, Shtrikman, Mahalu,
  Abusch-Magder, Meirav, and Kastner}}]{Goldhaber98}
\bibinfo{author}{\bibfnamefont{D.}~\bibnamefont{Goldhaber-Gordon}},
  \bibinfo{author}{\bibfnamefont{H.}~\bibnamefont{Shtrikman}},
  \bibinfo{author}{\bibfnamefont{D.}~\bibnamefont{Mahalu}},
  \bibinfo{author}{\bibfnamefont{D.}~\bibnamefont{Abusch-Magder}},
  \bibinfo{author}{\bibfnamefont{U.}~\bibnamefont{Meirav}}, \bibnamefont{and}
  \bibinfo{author}{\bibfnamefont{M.~A.} \bibnamefont{Kastner}},
  \bibinfo{journal}{Nature} \textbf{\bibinfo{volume}{391}},
  \bibinfo{pages}{156} (\bibinfo{year}{1998}).

\bibitem[{\citenamefont{Hentschel and Guinea}(2007)}]{Hentschel_2007}
\bibinfo{author}{\bibfnamefont{M.}~\bibnamefont{Hentschel}} \bibnamefont{and}
  \bibinfo{author}{\bibfnamefont{F.}~\bibnamefont{Guinea}},
  \bibinfo{journal}{Phys. Rev. B} \textbf{\bibinfo{volume}{76}},
  \bibinfo{pages}{115407} (\bibinfo{year}{2007}).

\bibitem[{\citenamefont{Neto et~al.}(2009{\natexlab{b}})\citenamefont{Neto,
  Kotov, Nilsson, Pereira, Peres, and Uchoa}}]{CastroNeto2009}
\bibinfo{author}{\bibfnamefont{A.~C.} \bibnamefont{Neto}},
  \bibinfo{author}{\bibfnamefont{V.}~\bibnamefont{Kotov}},
  \bibinfo{author}{\bibfnamefont{J.}~\bibnamefont{Nilsson}},
  \bibinfo{author}{\bibfnamefont{V.}~\bibnamefont{Pereira}},
  \bibinfo{author}{\bibfnamefont{N.}~\bibnamefont{Peres}}, \bibnamefont{and}
  \bibinfo{author}{\bibfnamefont{B.}~\bibnamefont{Uchoa}},
  \bibinfo{journal}{Solid State Communications} \textbf{\bibinfo{volume}{149}}
  (\bibinfo{year}{2009}{\natexlab{b}}).

\bibitem[{\citenamefont{Uchoa et~al.}(2009)\citenamefont{Uchoa, Yang, Tsai,
  Peres, and Neto}}]{Uchoa_2009}
\bibinfo{author}{\bibfnamefont{B.}~\bibnamefont{Uchoa}},
  \bibinfo{author}{\bibfnamefont{L.}~\bibnamefont{Yang}},
  \bibinfo{author}{\bibfnamefont{S.-W.} \bibnamefont{Tsai}},
  \bibinfo{author}{\bibfnamefont{N.~M.~R.} \bibnamefont{Peres}},
  \bibnamefont{and} \bibinfo{author}{\bibfnamefont{A.~H.~C.}
  \bibnamefont{Neto}}, \bibinfo{journal}{Phys. Rev. Lett.}
  \textbf{\bibinfo{volume}{103}} (\bibinfo{year}{2009}).

\bibitem[{\citenamefont{Vojta et~al.}(2010)\citenamefont{Vojta, Fritz, and
  Bulla}}]{Vojta_2010}
\bibinfo{author}{\bibfnamefont{M.}~\bibnamefont{Vojta}},
  \bibinfo{author}{\bibfnamefont{L.}~\bibnamefont{Fritz}}, \bibnamefont{and}
  \bibinfo{author}{\bibfnamefont{R.}~\bibnamefont{Bulla}},
  \bibinfo{journal}{Eurphys. Lett.} \textbf{\bibinfo{volume}{90}},
  \bibinfo{pages}{27006} (\bibinfo{year}{2010}).

\bibitem[{\citenamefont{Zhu and Berakdar}(2011)}]{Zhu_2011}
\bibinfo{author}{\bibfnamefont{Z.-G.} \bibnamefont{Zhu}} \bibnamefont{and}
  \bibinfo{author}{\bibfnamefont{J.}~\bibnamefont{Berakdar}},
  \bibinfo{journal}{Phys. Rev. B} \textbf{\bibinfo{volume}{84}}
  (\bibinfo{year}{2011}).

\bibitem[{\citenamefont{Saha et~al.}(2010)\citenamefont{Saha, Paul, and
  Sengupta}}]{Sengupta_2011}
\bibinfo{author}{\bibfnamefont{K.}~\bibnamefont{Saha}},
  \bibinfo{author}{\bibfnamefont{I.}~\bibnamefont{Paul}}, \bibnamefont{and}
  \bibinfo{author}{\bibfnamefont{K.}~\bibnamefont{Sengupta}},
  \bibinfo{journal}{Phys. Rev. B} \textbf{\bibinfo{volume}{81}},
  \bibinfo{pages}{165446} (\bibinfo{year}{2010}).

\bibitem[{\citenamefont{Chao and Aji}(2011)}]{Chao_2011}
\bibinfo{author}{\bibfnamefont{S.-P.} \bibnamefont{Chao}} \bibnamefont{and}
  \bibinfo{author}{\bibfnamefont{V.}~\bibnamefont{Aji}},
  \bibinfo{journal}{Phys. Rev. B} \textbf{\bibinfo{volume}{83}}
  (\bibinfo{year}{2011}).

\bibitem[{\citenamefont{Kharitonov and Kotliar}(2013)}]{Kharitonov_2013}
\bibinfo{author}{\bibfnamefont{M.}~\bibnamefont{Kharitonov}} \bibnamefont{and}
  \bibinfo{author}{\bibfnamefont{G.}~\bibnamefont{Kotliar}},
  \bibinfo{journal}{Phys. Rev. B} \textbf{\bibinfo{volume}{88}}
  (\bibinfo{year}{2013}).

\bibitem[{\citenamefont{Mitchell and Fritz}(2013)}]{Mitchell_2013}
\bibinfo{author}{\bibfnamefont{A.~K.} \bibnamefont{Mitchell}} \bibnamefont{and}
  \bibinfo{author}{\bibfnamefont{L.}~\bibnamefont{Fritz}},
  \bibinfo{journal}{Phys. Rev. B} \textbf{\bibinfo{volume}{88}}
  (\bibinfo{year}{2013}).

\bibitem[{\citenamefont{Fritz and Vojta}(2013)}]{Fritz_2013}
\bibinfo{author}{\bibfnamefont{L.}~\bibnamefont{Fritz}} \bibnamefont{and}
  \bibinfo{author}{\bibfnamefont{M.}~\bibnamefont{Vojta}},
  \bibinfo{journal}{Rep. Prog. Phys.} \textbf{\bibinfo{volume}{76}},
  \bibinfo{pages}{032501} (\bibinfo{year}{2013}).

\bibitem[{\citenamefont{Glossop et~al.}(2005)\citenamefont{Glossop, Jones, and
  Logan}}]{Glossop_2005}
\bibinfo{author}{\bibfnamefont{M.~T.} \bibnamefont{Glossop}},
  \bibinfo{author}{\bibfnamefont{G.~E.} \bibnamefont{Jones}}, \bibnamefont{and}
  \bibinfo{author}{\bibfnamefont{D.~E.} \bibnamefont{Logan}},
  \bibinfo{journal}{J. Phys. Chem. B} \textbf{\bibinfo{volume}{109}},
  \bibinfo{pages}{6564} (\bibinfo{year}{2005}).

\bibitem[{\citenamefont{Nazarov and Blanter}(2009)}]{Nazarov_2009}
\bibinfo{author}{\bibfnamefont{Y.~V.} \bibnamefont{Nazarov}} \bibnamefont{and}
  \bibinfo{author}{\bibfnamefont{Y.~M.} \bibnamefont{Blanter}}, in
  \emph{\bibinfo{booktitle}{Quantum Transport}} (\bibinfo{publisher}{Cambridge
  University Press}, \bibinfo{year}{2009}).

\bibitem[{\citenamefont{Liang et~al.}(2002)\citenamefont{Liang, Bockrath, and
  Park}}]{Liang_2002}
\bibinfo{author}{\bibfnamefont{W.}~\bibnamefont{Liang}},
  \bibinfo{author}{\bibfnamefont{M.}~\bibnamefont{Bockrath}}, \bibnamefont{and}
  \bibinfo{author}{\bibfnamefont{H.}~\bibnamefont{Park}},
  \bibinfo{journal}{Phys. Rev. Lett.} \textbf{\bibinfo{volume}{88}},
  \bibinfo{pages}{126801} (\bibinfo{year}{2002}).

\bibitem[{\citenamefont{Balandin}(2011)}]{Balandin_2011}
\bibinfo{author}{\bibfnamefont{A.~A.} \bibnamefont{Balandin}},
  \bibinfo{journal}{Nature Materials} \textbf{\bibinfo{volume}{10}},
  \bibinfo{pages}{569} (\bibinfo{year}{2011}).

\bibitem[{\citenamefont{Dubi and Ventra}(2011)}]{Dubi2011}
\bibinfo{author}{\bibfnamefont{Y.}~\bibnamefont{Dubi}} \bibnamefont{and}
  \bibinfo{author}{\bibfnamefont{M.~D.} \bibnamefont{Ventra}},
  \bibinfo{journal}{Rev. Mod. Phys.} \textbf{\bibinfo{volume}{83}},
  \bibinfo{pages}{131} (\bibinfo{year}{2011}).

\bibitem[{\citenamefont{Zubarev}(1960)}]{Zubarev}
\bibinfo{author}{\bibfnamefont{D.}~\bibnamefont{Zubarev}},
  \bibinfo{journal}{Usp. Fiz. Nauk} \textbf{\bibinfo{volume}{71}},
  \bibinfo{pages}{71} (\bibinfo{year}{1960}).

\bibitem[{\citenamefont{Lacroix}(1981)}]{Lacroix98}
\bibinfo{author}{\bibfnamefont{C.}~\bibnamefont{Lacroix}},
  \bibinfo{journal}{Journal of Physics F: Metal Physics}
  \textbf{\bibinfo{volume}{11}}, \bibinfo{pages}{2389} (\bibinfo{year}{1981}).

\bibitem[{\citenamefont{Meir et~al.}(1993)\citenamefont{Meir, Wingreen, and
  Lee}}]{Meir1993}
\bibinfo{author}{\bibfnamefont{Y.}~\bibnamefont{Meir}},
  \bibinfo{author}{\bibfnamefont{N.~S.} \bibnamefont{Wingreen}},
  \bibnamefont{and} \bibinfo{author}{\bibfnamefont{P.~A.} \bibnamefont{Lee}},
  \bibinfo{journal}{Phys. Rev. Lett.} \textbf{\bibinfo{volume}{70}},
  \bibinfo{pages}{2601} (\bibinfo{year}{1993}).

\bibitem[{\citenamefont{Kashcheyevs et~al.}(2006)\citenamefont{Kashcheyevs,
  Aharony, and Entin-Wohlman}}]{Kashcheyevs2006}
\bibinfo{author}{\bibfnamefont{V.}~\bibnamefont{Kashcheyevs}},
  \bibinfo{author}{\bibfnamefont{A.}~\bibnamefont{Aharony}}, \bibnamefont{and}
  \bibinfo{author}{\bibfnamefont{O.}~\bibnamefont{Entin-Wohlman}},
  \bibinfo{journal}{Phys. Rev. B} \textbf{\bibinfo{volume}{73}},
  \bibinfo{pages}{125338} (\bibinfo{year}{2006}).

\bibitem[{\citenamefont{Glossop and Logan}(2003)}]{Glossop_2003}
\bibinfo{author}{\bibfnamefont{M.~T.} \bibnamefont{Glossop}} \bibnamefont{and}
  \bibinfo{author}{\bibfnamefont{D.~E.} \bibnamefont{Logan}},
  \bibinfo{journal}{Europhysics Lett.} \textbf{\bibinfo{volume}{61}},
  \bibinfo{pages}{810} (\bibinfo{year}{2003}).

\bibitem[{\citenamefont{Onsager}(1931)}]{onsager_1931}
\bibinfo{author}{\bibfnamefont{L.}~\bibnamefont{Onsager}},
  \bibinfo{journal}{Phys. Rev.} \textbf{\bibinfo{volume}{37}},
  \bibinfo{pages}{405} (\bibinfo{year}{1931}).

\bibitem[{\citenamefont{Mahan}(1990)}]{mahan1990many}
\bibinfo{author}{\bibfnamefont{G.}~\bibnamefont{Mahan}},
  \emph{\bibinfo{title}{Many-Particle Physics}} (\bibinfo{publisher}{Springer
  US}, \bibinfo{year}{1990}).

\bibitem[{\citenamefont{Jie~Ren et~al.}(2012)\citenamefont{Jie~Ren, Gubernatis,
  and Chen~Wang}}]{PhysRevB.85.155443}
\bibinfo{author}{\bibfnamefont{J.-X.~Z.} \bibnamefont{Jie~Ren}},
  \bibinfo{author}{\bibfnamefont{J.~E.} \bibnamefont{Gubernatis}},
  \bibnamefont{and} \bibinfo{author}{\bibfnamefont{B.~L.}
  \bibnamefont{Chen~Wang}}, \bibinfo{journal}{Phys. Rev. B}
  \textbf{\bibinfo{volume}{85}}, \bibinfo{pages}{155443}
  (\bibinfo{year}{2012}).

\bibitem[{\citenamefont{Wu et~al.}(2005)\citenamefont{Wu, Cao, and
  Ahn}}]{Fano1}
\bibinfo{author}{\bibfnamefont{B.~H.} \bibnamefont{Wu}},
  \bibinfo{author}{\bibfnamefont{J.~C.} \bibnamefont{Cao}}, \bibnamefont{and}
  \bibinfo{author}{\bibfnamefont{K.-H.} \bibnamefont{Ahn}},
  \bibinfo{journal}{Phys. Rev. B} \textbf{\bibinfo{volume}{72}},
  \bibinfo{pages}{165313} (\bibinfo{year}{2005}).

\bibitem[{\citenamefont{Liu and Yang}(2010)}]{Fano2}
\bibinfo{author}{\bibfnamefont{Y.~S.} \bibnamefont{Liu}} \bibnamefont{and}
  \bibinfo{author}{\bibfnamefont{X.~F.} \bibnamefont{Yang}},
  \bibinfo{journal}{Journal of App. Phys. Lett} \textbf{\bibinfo{volume}{108}}
  (\bibinfo{year}{2010}).

\bibitem[{\citenamefont{Sevincli et~al.}(2011)\citenamefont{Sevincli, Sevik,
  Ca\u{g}in, and Cuniberti}}]{kphonon}
\bibinfo{author}{\bibfnamefont{H.}~\bibnamefont{Sevincli}},
  \bibinfo{author}{\bibfnamefont{C.}~\bibnamefont{Sevik}},
  \bibinfo{author}{\bibfnamefont{T.}~\bibnamefont{Ca\u{g}in}},
  \bibnamefont{and}
  \bibinfo{author}{\bibfnamefont{G.}~\bibnamefont{Cuniberti}},
  \bibinfo{journal}{Science Reports} \textbf{\bibinfo{volume}{3}},
  \bibinfo{pages}{035405} (\bibinfo{year}{2011}).

\end{thebibliography}

\end{document}